\documentclass[aps, singlecolumn, showpacs]{revtex4-2}
\usepackage{amsmath}
\begin{document}
\title{Physical mechanism for the geometric phases in optics and angular momentum holonomy}
\author{S C Tiwari}
\affiliation{Department of Physics, Institute of Science, Banaras Hindu University, Varanasi 221005, and \\ Institute of Natural Philosophy \\
Varanasi India\\ Email address: $vns\_sctiwari@yahoo.com$ \\}
\begin{abstract}
Vast literature on the experiments and mathematical formulations on the geometric phases signifies the importance of this subject. Physical mechanism for the origin of the geometric phases in optics was suggested in 1992 by the author in terms of the exchange of the angular momentum. Some of the literature has taken notice of it, however, the real import of the suggested mechanism and the angular momentum holonomy conjecture has remained elusive. The present contribution delineates the prevailing confusions and offers a thorough discussion on the nature of the light waves to resolve the conceptual issues. It is argued that at a fundamental level there exist only two types of geometric phases arising from (i) the geometry of the wave vector space, and (ii) the polarization state space. For the light beams of definite polarization  spatial modes (HG or LG modes) the geometric phase has origin in the wave vector space that is different than the spin redirection phase for the polarized light. Angular momentum holonomy is proposed to be equivalent to the geometric phase in general.
\end{abstract}
\maketitle
\section{\bf Introduction}

Mathematical notion of holonomy as an underlying general concept in physics emerges from the introductory surveys and the reprinted papers in the volume published in 1989 \cite{1}. Angular momentum (AM) holonomy as a physical mechanism for the origin of the geometric phases (GP) in optics was proposed in 1992 \cite{2}. Apparently, this suggestion found support in the literature \cite{3,4,5}. However, it was argued \cite{6} that the import of this conjecture was not understood properly: the AM exchange between the light and the optical elements that occurs during the polarization state space or mode space transformations does not represent AM holonomy proposed in \cite{2}. van Enk and Nienhuis \cite{7} indicated a possible deeper connection between AM of light and geometric phases \cite{2}, however, the nature of this connection remained unclear. In a nice review on the Berry phase in 1996, Banerjee \cite{8} presents twin contributions of my paper \cite{2} correctly: physical insights gained on classical or quantum nature of GP in optics, and AM holonomy conjecture that involves the role of the electromagnetic (EM) vector potential. The assumed physical reality of the EM potential is crucial to define AM holonomy. Recently, the origin of the Aharonov-Bohm (AB) effect has been traced to the exchange of modular angular momentum associated with pure gauge field \cite{9}. Does AM holonomy conjecture have general validity for geometric and topological phases? Is AM holonomy measurable? The aim of the present paper is to seek answers to these questions.

Angular momentum of EM radiation fields and light beams known for more than a century continues to offer surprises and intrigues. Advances in technology to shape, control, and manipulate radiation field and light have led to interesting applications as well as the acceptance of so called counter-intuitive features associated with light and photon. Unfortunately, at a fundamental level there remain unresolved questions on the physical reality of photon, wave-particle attributes of light, reality of EM potential, and spin AM of photon. Textbooks, for example, \cite{10,11,12,13,14,15} and monographs \cite{16, 17, 18} offer nice illuminating discussion on some of these issues that somehow current literature does not acknowledge/appreciate. 

The paper is organized as follows. In the next section, we present  a thorough discussion on the physical concept of AM holonomy. Almost axiomatically, the literature on light waves and photons asserts that these are electromagnetic waves and EM field quanta respectively. However, in this framework there are a number of conceptual problems that are pointed out in Section III.  The meaning of pure vector waves, EM waves, structured light, and the single-photon states are re-visited in this section. Section IV is devoted to highlight the occurrence of holonomy in non-Euclidean geometry and the geometry of paths. The manifestation of geometrical holonomy in physical systems is discussed in Section V. A generalized principle is proposed that geometric phase is equivalent to AM holonomy.  New directions for testing AM holonomy are suggested. The last section summarizes the main results of the present work.

\section{\bf Angular Momentum Holonomy: Physical Concepts}

The AB effect \cite{19} and Pancharatnam phase \cite{20} are the widely discussed important examples of non-trivial geometric effects in physics. A remarkable anticipation of holonomy is that arising from a kinematic description of electron motion \cite{21}. This holonomy has been interpreted to seek the topological origin of spin of electron \cite{22, 23}. An over-view on geometry and topology relevant for AB effect and GP in optics given in \cite{1, 8, 24} will put several claims made in a recent review \cite{25} in the right perspective .

Early review on Pancharatnam phase and Rytov-Vladimirskii phase in optics  can be found in \cite{26}. The role of angular momentum exchange in GP in optics was considered in \cite{27} using a generalized Poincare sphere for angular momentum of light. The authors \cite{27} note that, 'it is angular momentum exchange of the light with the optical components which is the common origin of both phases'. Pancharatnam phase arises in the polarization state space of light geometrically described by a 2-sphere $S^2$, known as Poincare sphere. Spin redirection phase or Rytov-Vladimirskii-Chiao-Wu (RVCW) phase arises in the wave-vector state space of light geometrically represented by $S^2$ of directions in the momentum space. It is obvious that changing the polarization of light or the momentum of light needs respectively the transfer of spin or orbital AM of light to the optical elements. In the Beth experiment \cite{28} torque exerted by polarized light on optical elements gave direct measurement of spin AM of light. However, this kind of AM exchange that necessarily occurs in GP as pointed out in \cite{27} is not AM holonomy proposed in \cite{2}. We elaborate on this issue, and also present new aspects on the concept of AM holonomy.

The idea of AM holonomy in analogy to the role of the EM potentials in the AB effect was proposed in terms of the pure divergence term of AM density being a difference between the gauge-invariant and gauge-non-invariant AM tensor of the EM fields \cite{2}. The origin of the Pancharatnam phase was qualitatively related with spin AM exchange, and that of spin redirection phase was related with that of the orbital AM exchange. The crucial point is that both spin and orbital AM holonomies represent the constant levels of AM values that amount to torque-free case; and, it is these nontrivial constant shifts in the AM that are related with the geometric phases. When this paper was written \cite{2}, the light beams carrying orbital AM \cite{29} were not realized, and one of the objections raised by the reviewer to the proposed orbital AM exchange was based on Blatt and Weisskopf \cite{30}: the  AM of multipole radiation fields in the case of nuclear transitions was well known \cite{30}, however, the idea that a light beam possessed orbital AM was considered unacceptable. In fact, unlike energy and linear momentum associated with the light or EM radiation that found acceptance and experimental demonstration rather quickly, the AM and its mechanical effects took much longer and uncertain path. Though Poynting in 1909 \cite{31} envisaged AM of light it was only in 1935 that direct measurement was made by Beth \cite{28}. Light quantum as a packet of energy $h \nu$ for Planck oscillator suggested in 1905 was endowed with linear momentum of $h \nu/c$ in 1917 by Einstein \cite{18, 32}; however, even in 1926 Einstein maintained that photon angular momentum (spin) was less deep than linear momentum in quantum theory \cite{32}. In contrast, Bose had envisaged intrinsic photon spin as early as 1924 \cite{33}.  

The confusions on the physical meaning of AM of light and photon spin have not ended even today. Let us consider spin AM of  EM waves. Plane wave representation of light is a mathematical idealization that works nicely for pedagogy as well as for the description of fundamental physical concepts \cite{10, 11, 13, 16}. Assuming a special reference frame in which the momentum of the light field is directed along z-axis, the orbital AM is zero along this axis since $\hbar k_x =\hbar k_y =0$. Does there exist spin AM for plane waves? Simmons and Guttmann \cite{15} in section 9.1 give an illuminating discussion on the AM of approximate plane polarized waves such that the light beam has finite transverse extent within a cylinder about the z-axis. Plane wave representation for right circularly polarized (RCP) and left circularly polarized (LCP) light is modified such that 
\begin{equation}
{\bf E} = [E_0(r) (\hat{e}_r \pm i \hat{e}_\phi) + f(r) \hat{e}_z] e^{i(k z -\omega t \pm \phi \mp \pi /4)}
\end{equation}
Using the Poynting vector the ratio of the calculated z-component of AM per unit length to energy per unit length for arbitrary function $f(r)$ is obtained to be $\pm \frac{1}{\omega}$. Authors \cite{15} remark that this result is 'consistent with the quantum viewpoint of photons each  carrying energy $\hbar \omega$ and angular momentum $\pm \hbar$'.

In the language of quantized radiation field, the wave vector ${\bf k}$ and the polarization index $s$ define the modes of the plane wave representation. In the basis of circular polarization, it can be shown that the spin AM associated with any ${\bf k}$ is 
\begin{equation}
{\bf S} =\frac{ \hbar {\bf k}}{|{\bf k}|} (N_r -N_l)
\end{equation}
where $N_r, N_l$ are the number of RCP and LCP photons respectively \cite{14,18}.
 
The profound significance of the orbital AM of light beams explored in \cite{29} becomes obvious in the backdrop of briefly noted historical developments above, and the then prevailing mainstream thinking on AM in quantum optics \cite{15, 18} and photon spin in QED \cite{17}. The existence of phase singularities for the laser beams was already known and had been experimentally observed \cite{34,35}, however, the connection between phase vortices and orbital AM was first made by Allen et al \cite{29}. A Laguerre-Gaussian (LG) laser mode of a cylindrical light beam in the paraxial approximation having azimuthal dependence $e^{i l \phi}$ is shown to possess orbital AM along z-axis such that the ratio of the AM flux to that of the energy calculated from the Poynting vector and the energy density of EM fields is equal to $l/\omega$, and ratio of orbital to linear AM is $l/k_z$. If the laser mode is described in terms of the product of Hermite polynomials $H_n(x) H_m(y)$, i. e. HG-modes, then the associated orbital AM is zero. Introducing a $\pi /2$ phase change in the HG-modes converts them to LG-modes; Tamm and Weiss \cite{34} had achieved required phase changes for the conversion of first-order modes. 

It has to be emphasized that analogy to quantum mechanical interpretation attributing orbital AM $l \hbar$ per photon \cite{29}, and spin AM $\pm \hbar$ for photon \cite{15} could become misleading since the word photon used here does not represent a single physical object or a gauge boson or a quantized massless vector field. In section 9.8 of \cite{10}, for a multipole radiation field of order $(l, m)$ it is stated that it has $m\hbar$ units of z-component of AM per photon of energy $\hbar \omega$. However, classical and quantum values of the squared quantities differ, and in the later part of section 9.8, this difference is explained invoking uncertainty principle. In section 1.1 \cite{10} a short informative discussion is worth reading: discrete photon versus continuous fields, and the validity limit of semiclassical treatment for interaction of light with matter. To differentiate a physically real photon and a single photon state used in quantum optics we use the term "topological photon" for the former \cite{36, 37}.

The measurement of the mechanical torque due to the transfer of orbital AM from the light beams to a particle or an optical element was suggested to prove the existence of orbital AM of LG-modes \cite{29} in analogy to spin AM transfer from polarized light in the Beth experiment. Direct observation of this effect in 1995 \cite{38} left no doubt on the physical concept of orbital AM carrying light beams proposed in \cite{29}. Rapidly, enormous activities in this field of optics followed world-over. One of the interesting studies of concern here was on the geometric phase associated with the cyclic changes in the transverse spatial modes of Gaussian light beams \cite{3}. A direct measurement of this geometric phase was reported in 2003 \cite{39}. Let us denote LG-modes by $LG^l_p$ where $l,p$ are azimuthal and radial indices respectively, and HG-modes by $HG_{nm}$. In the experiment, $LG^{-1}_0$ to $HG_{01}$, a rotation, and finally to the initial mode $LG^{-1}_0$ completes the cycle in the mode-space. The exchange of AM in this kind of GP was emphasized in \cite{3,39}. There arise two questions: Is it a new GP in optics? Does the transfer of AM discussed in these papers represent the AM holonomy conjecture \cite{2}?

In a detailed elucidation of AM holonomy \cite{6} and a critical appraisal on \cite{3,39} it was concluded that (i) the GP in transverse mode space was not new, and at a fundamental level this phase occurs as a special case of GP in momentum space, and (ii) the AM transfer discussed in connection with the cycles in the mode space corresponds to the standard AM transfer that is necessarily accompanied when a polarization state is changed, as the one that occurs in the case of Beth experiment \cite{28}, or, the orbital AM transfer measured in \cite{38}. It is clear that the AM transfer considered in \cite{3,39} is not AM holonomy. A recent review \cite{25} repeats the claim that GP in mode space is a new GP of light, and misinterprets AM holonomy suggested in \cite{2} to the AM transfer proposed in \cite{27}. It has to be emphasized that the AM transfer discussed in \cite{3,27,39} is the usual one akin to Beth experiment \cite{28} or orbital AM transfer \cite{38} at the intermediate stages of cyclic evolution on the state-space of light. Let us quote from \cite{25}: "This indicates that geometric phases are mediated by a variation of orbital angular momentum, in the same way that polarization transformations that generate a PB phase involve variation of the spin angular momentum." Here PB is an abbreviation for Pancharatnam-Berry. However, in our view GP in the polarization state space should be termed just Pancharatnam phase: viewed in conjunction with Berry phase unnecessary formal artefacts introduce obscurity. Morevover, Berry himself does not quite agree with the terminology Pancharatnam-Berry phase (private communication). In spite of \cite{6} and a later exposition \cite{40}, the lack of clarity on the nature of GP in mode space, and obscurity on AM holonomy conjecture persists. We believe resolution of this issue is of great importance. To address this question we proceed in three steps: the nature and physical attributes of vector waves, mathematical aspects of GP, and thorough treatment of AM holonomy.

\section{\bf Light and vector waves}

Prior to the Maxwell's electromagnetic wave theory of light a large number of optical phenomena had been discovered and understood based on geometrical optics and scalar wave optics \cite{11}. Chapter 10 for spherical scalar waves and Chapter 12 for scalar wave diffraction in \cite{13} may also be useful. In a remarkable pedagogical approach Simmons and Guttmann \cite{15} discuss physical properties of light based solely on the various kinds of measurements using light and optical systems in the first four chapters.  It has become almost axiomatic to state that light waves are vector waves. However, there are subtle aspects related with the nature of the vector waves that need careful consideration. One of the important issues is that two vector fields, namely, electric and magnetic fields are needed for the description of light. Another question is that of the plane wave representation. Most of the textbooks discuss these issues with varied degree of emphasis and outlook. Unfortunately, in the current literature some authors tend to ignore or misinterpret basic concepts to claim so called new counter-intuitive properties of light. Added emphasis on basic concepts and a new perspective motivate re-visiting light waves here. 

A concise and reasonably rigorous presentation on both scalar and vector waves is given in the first chapter and the wave equation for EM potentials is discussed in the second chapter of Born and Wolf \cite{11}. For the bounded EM waves, for example, in waveguides and cavities, and also for the radiating systems like antennas, it is convenient to classify the vector waves in terms of elementary types: (i) transverse EM (TEM) waves having zero electric and magnetic fields in the direction of propagation, i. e. both fields lie in the transverse plane, (ii) transverse electric (TE) waves having only magnetic field in the direction of propagation; also termed as H-waves, and (iii) transverse magnetic (TM) waves having no magnetic field along the direction of propagation; also known as E-waves. In actual systems one may have to use hybrid-modes for the EM waves. We refer to chapter 8 in \cite{10}, and for more detailed nice discussion to \cite{12,13}.

{\bf SCALAR WAVES}

Scalar waves for the solution of the homogeneous wave equation have simplest representation in the form of plane waves. The field equation
\begin{equation}
\nabla^2 \Phi -\frac{1}{c^2} \frac{\partial^2 \Phi}{\partial t^2}=0
\end{equation}
has a solution
\begin{equation}
\Phi({\bf r},t) =\Phi_0 \cos ({\bf k}\cdot {\bf r} -\omega t +\delta)
\end{equation}
The propagation vector ${\bf k}$ and angular frequency $\omega=2\pi \nu$ are related by 
\begin{equation}
|{\bf k}|^2 =\omega^2/c^2
\end{equation}
$\Phi_0$ is a constant field amplitude, and $\delta$ is a constant phase. Fixed frequency $\nu$ means we are considering monochromatic plane wave. 

A generalization to the solution (4) is a monochromatic inhomogeneous scalar wave
\begin{equation}
\Phi = \Phi_0({\bf r}) \cos (\omega t - g({\bf r}))
\end{equation}
In this solution, the wave surfaces or co-phasal surfaces defined by 
\begin{equation}
g({\bf r})= constant
\end{equation}
may not always coincide with the surfaces of constant amplitude. If $d(\omega t-g({\bf r})) =0$ then the phase velocity, i. e. the speed of the co-phasal surfaces is defined to be 
\begin{equation}
v_p({\bf r})=\frac{\omega}{{|\bf \nabla} g|}
\end{equation}

Note that physical fields are real, and all measurable physical quantities are also real. However, it is often convenient to use complex representation for the ease of calculations. The scalar wave (5) could be re-written in a complex form
\begin{equation}
\Phi = \Phi_0 e^{i(g({\bf r}) -\omega t)}
\end{equation}
In contrast to the real scalar field wave equation  (3), the Scr$\ddot{o}$dinger wave equation for the wavefunction in quantum theory is intrinsically complex
\begin{equation}
-\frac{\hbar^2}{2m} \nabla^2 \Psi = i \hbar \frac{\partial \Psi}{\partial t}
\end{equation}
Originally, Schr$\ddot{o}$dinger interpreted complex scalar field $\Psi$ in Eq.(10) to represent electrically charged electron; however, Born's probabilistic interpretation, Heisenberg's uncertainty relation, and wave-particle duality ultimately led to the development of the orthodox or the Copenhagen interpretation of quantum mechanics. The physical reality of the Schr$\ddot{o}$dinger wavefunction continues to remain an unsettled issue among physicists; the presence of the imaginary unit $i$ in the wave equation (10) is believed to be one of the reasons for the mysterious quantum world \cite{41}.

Relativistic field theory based on the variational principle of the action functional offers new insights on the symmetries and conservation laws in addition to the derivation of the field equations as Euler-Lagrange equations of motion. For a real scalar field the Lagrangian density is
\begin{equation}
\mathcal{L}_{scalar} = \partial^\mu \Phi \partial_\mu \Phi
\end{equation}
and the action integral is
\begin{equation}
S_{scalar} = \int_\Omega \mathcal{L}_{scalar} d^4x
\end{equation}
Variational principle treating $\Phi$ as independent field variable leads to the field equation (3). Invariance of the Lagrangian density is equivalent to that of the action integral. The integration region $\Omega$ is the volume between two surfaces of constant time $t$, and the integral is Lorentz invariant. To assume a specific frame of reference is, however, not a manifest Lorentz invariant concept. Instead of the surface at a constant $t$, Weiss \cite{42} suggested an invariant concept of a space-like surface \cite{17}. 

Action (12) is invariant under infinitesimal space-time translations
\begin{equation}
x^\mu ~\rightarrow ~ x^\mu + \delta x^\mu
\end{equation}
The corresponding conserved quantity is the canonical energy-momentum tensor
\begin{equation}
E^{\mu\nu}_{scalar} = \partial^\mu \Phi \partial^\nu \Phi - \eta^{\nu\mu} \mathcal{L}_{scalar}
\end{equation}
One also has the invariance under infinitesimal Lorentz transformations
\begin{equation}
x^\mu ~ \rightarrow ~ x^\mu + \omega^{\mu\nu} x_\nu
\end{equation}
resulting into a conserved quantity represented by a third rank tensor for the angular momentum density. For the scalar field action (12) we have
\begin{equation}
J^{\mu\nu\lambda} = (E^{\mu\nu}_{scalar} x^\lambda -E^{\mu\lambda}_{scalar} x^\nu)
\end{equation}
The AM density of the field is $M^{\nu\lambda} =J^{0\nu\lambda}$. In fact, the space-space components $M^{ij}$ define orbital AM of the scalar field.

{\bf VECTOR WAVES}

A straightforward generalization of the wave equation (3) for a 4-vector field $V^\mu$ is
\begin{equation}
\nabla^2 V^\mu -\frac{1}{c^2} \frac{\partial^2 V^\mu}{\partial t^2}=0
\end{equation}
For the vector field ${\bf V}({\bf r}, t)$ the plane wave solution of Eq.(17) is 
\begin{equation}
{\bf V}({\bf r}, t) = {\bf V}_0 e^{i(k^\mu x_\mu +\delta)}
\end{equation}
The vector nature of the wave is embodied in the constant vector ${\bf V}_0$. Similar to the solution (6) for the scalar waves one obtains the inhomogeneous vector waves for the vector field; in the cartesian coordinate system monochromatic vector waves could be represented by
\begin{equation}
{\bf V}({\bf r}, t) = V_x({\bf r} ) \cos (\omega t - g_1({\bf r})) \hat{e}_x +V_y({\bf r} ) \cos (\omega t - g_2({\bf r})) \hat{e}_y +V_z({\bf r} ) \cos (\omega t - g_3({\bf r})) \hat{e}_z
\end{equation}

Constant vector ${\bf V}_0$ in Eq.(18) defines the polarization of the plane vector waves. In the framework of the relativistic field theory, Eq.(17) represents a massless vector field: considerations of Wigner's little group, transversality, and the helicity states would acquire significance in this case \cite{17}.

Inhomogeneous monochromatic vector waves (19) discussed in detail in Section 1.4.3 of \cite{11} represent the class of vector waves that in recent years have invited world-wide attention in connection with space-variant polarization, polarization singularities, and vector vortex beams of light. To set the current research on the vector waves of light in right perspective a brief account on the nature of vector waves following \cite{11} is given here. 

For a fixed frequency the complex representation of the vector field is 
\begin{equation}
{\bf V}({\bf r}, t) = ({\bf p}({\bf r}) +i {\bf q}({\bf r})) e^{- i \omega t}
\end{equation}
The meaning of polarization and the direction of propagation of the vector waves become highly nontrivial issues as both phase and amplitute are the functions of the space coordinates ${\bf r}$. Choosing a point ${\bf r} ={\bf r}_0$ the behaviour of the vector field can be described specifying ${\bf p}({\bf r}_0),~{\bf q}({\bf r}_0)$. Introducing a scalar function $\epsilon$
\begin{equation}
{\bf p}({\bf r}) +i {\bf q}({\bf r})= ({\bf a} + i{\bf b}) e^{i \epsilon}
\end{equation}
the orthogonality of $({\bf a},~{\bf b})$ could be ensured if 
\begin{equation}
\tan 2 \epsilon = \frac{2 {\bf p}\cdot {\bf  q}}{ {\bf p}^2 -{\bf q}^2}
\end{equation}
without loss of generality one can assume $|{\bf a}| \geq |{\bf b}|$.

In cartesian coordinate system, let ${\bf r}_0$ be the origin, then ${\bf V}$ given by (19) can be re-written as 
\begin{equation}
V_x = a \cos (\omega t -\epsilon), ~ V_y = b \sin (\omega t -\epsilon), ~ V_z = 0
\end{equation}
Eq.(23) defines a polarization ellipse
\begin{equation}
\frac{V_x^2}{a^2} + \frac{V_y^2}{b^2} =1
\end{equation}
In general, $({\bf p}, {\bf q}, {\bf \nabla} \epsilon)$ or $({\bf a}, {\bf b}, {\bf \nabla} \epsilon)$ determine the polarization and the propagation of the vector waves. The sign of the scalar product $(({\bf a} \times {\bf b} )\cdot {\bf \nabla} \epsilon)$ determines the sense of polarization (left or right). Polarization ellipse degenerates to circular polarization when $\epsilon$ is indeterminate, i. e. from Eq.(22)
\begin{equation}
2 {\bf p}\cdot {\bf  q}=  {\bf p}^2 -{\bf q}^2=0
\end{equation}
and to a straight line when $b^2=0$ or
\begin{equation}
p^2 q^2 =({\bf p}\cdot {\bf q})^2
\end{equation}
In this case the vector wave is linearly polarized.

An important remark from Born and Wolf \cite{11} is reproduced for the sake of clarity regarding the inhomogeneous vector waves: "Finally we stress that the term polarization refers to the behaviour at a particular point in the field, and that the state of polarization will therefore in general be different at different points of the field."

{\bf ELECTROMAGNETIC WAVES}

Source-free Maxwell field equations in vacuum for the vector fields ${\bf E}$ and ${\bf B}$ are the set of following four equations 
\begin{equation}
{\bf \nabla}\cdot {\bf E}=0
\end{equation}
\begin{equation}
{\bf \nabla}\cdot {\bf B}=0
\end{equation}
\begin{equation}
{\bf \nabla} \times {\bf E} =-\frac{1}{c} \frac{\partial {\bf B}}{\partial t}
\end{equation}
\begin{equation}
{\bf \nabla} \times {\bf B} =\frac{1}{c} \frac{\partial {\bf E}}{\partial t}
\end{equation}
Curl of (28) using (27) and (30) gives the wave equation for the electric field
\begin{equation}
\nabla^2 {\bf E} -\frac{1}{c^2} \frac{\partial^2 {\bf E}}{\partial t^2}=0
\end{equation}
Similarly, the curl of (30) using (28) and (29) gives the wave equation for the magnetic field 
\begin{equation}
\nabla^2 {\bf B} -\frac{1}{c^2} \frac{\partial^2 {\bf B}}{\partial t^2}=0
\end{equation}

Unlike the vector waves for the vector field ${\bf V}$ the EM waves comprise of two vector waves for the wave equations (31) and (32). Moreover, two of the four equations, (27) and (28) signify the transversality constraints for the electric and magnetic vector waves respectively. For a plane EM wave 
\begin{equation}
{\bf E} ={\bf E}_0 e^{i({\bf k}\cdot {\bf r} -\omega t)}
\end{equation}
\begin{equation}
{\bf B} ={\bf B}_0 e^{i({\bf k}\cdot {\bf r} -\omega t)}
\end{equation} 
substitution of (33) and (34) in the divergence equations (27)-(28) gives
\begin{equation}
{\bf k}\cdot {\bf E}_0 ={\bf k}\cdot {\bf B}_0=0
\end{equation}
The curl equation (29) shows that
\begin{equation}
{\bf B}_0={\bf k} \times {\bf E}_0
\end{equation}

Note that the constant amplitudes ${\bf E}_0, {\bf B}_0$ written in complex form provide simple explanation for the polarization of plane EM waves. Light polarization is usually described using complex electric field amplitude for historical reasons. Electric field vector is termed light vector. Remarks on aspects of terminology may be of interest given on page 28 of \cite{11}.

In some of the recent literatue on light beams/quantum optics much ado about complex quantitiesis is being made. To get rid off it let us emphasize that the Maxwell field equations founded on experimental laws describe real vector fields $({\bf E}, {\bf B})$, and complex representation is nothing more than a convenient choice for the calculational purposes. However, a cautionary remark on handling the complex representation on page 17 of \cite{11} is worth remembering. For example, the calculation of the time-averaged Poynting vector given in section 5.4 of \cite{13}. Complex propagation vector is introduced to take into account the wave propagation in a conducting medium or lossy diekectric materials phenomenologically. In the engineering applications attenuation and circuit losses are described introducing complex impedance and admittance \cite{12}.

A formal system of complex vector field equations equivalent to the Maxwell field equations has also been of interest \cite{43}. Define a complex vector field 
\begin{equation}
{\bf \Psi} = {\bf E} +i {\bf B}
\end{equation}
The wave equations (31) and (32) are combined to obtain an equivalent wave equation for the complex vector field ${\bf \Psi}$
\begin{equation}
\nabla^2 {\bf \Psi} -\frac{1}{c^2} \frac{\partial^2 {\bf \Psi}}{\partial t^2}=0
\end{equation}
The first order partial differential equations (29) and (30) are combined to derive field equation for ${\bf \Psi}$. Defining 
\begin{equation}
(S_i)_{jk}= i \epsilon_{ijk}
\end{equation}
and noting that ${\bf \nabla} \times {\bf E}~ \rightarrow~ \epsilon_{ijk} \frac{\partial E_k}{\partial x_j}$ the curl equations lead to
\begin{equation}
-c {\bf S}\cdot {\bf p} {\bf \Psi}= i \hbar \frac{\partial {\bf \Psi}}{\partial t}
\end{equation}
Here ${\bf p} = -i \hbar {\bf \nabla}$. The divergence equations (27)-(28) become
\begin{equation}
\frac{\partial \Psi_i}{\partial x_i}=0
\end{equation}
Quantum mechanical interpretation of Eq.(40) treating it as a Schr$\ddot{o}$dinger look-alike equation for photon wavefunction ${\bf \Psi}$ is conceptually flawed: photon is a massless particle, and the construction of Hilbert space for ${\bf \Psi}$ has no well-defined and acceptable procedure. The form of Eq.(40) resembles the Weyl equation for massless neutrino \cite{44}; however, quantized field theory of photon based on Eq.(40) does not exist. 

EM waves have two vector fields whose behaviour under discrete symmetry transformations is different \cite{10}. A true vactor under space inversion ${\bf r} \rightarrow -{\bf r}$ changes sign while a true scalar remains invariant. Pseudo-vector or axial vector does not change sign under space inversion while a pseudo-scalar changes sign. Assuming that the charge density is a scalar quantity the divergence equation for ${\bf E}$ in the presence of sources implies that ${\bf E}$ is a true vector. Now, the curl equation (30) shows that ${\bf B}$ is a pseudo-vector. Thus, though both ${\bf E}$ and ${\bf B}$ transform as vectors under the rotation group their nature under space inversion and time reversal is different. 

True vector field  description is possible for EM waves introducing EM potentials. From Eq.(28) one has
\begin{equation}
{\bf B} = {\bf \nabla} \times {\bf A}
\end{equation}
Making use of (42) the curl equation (29) implies that
\begin{equation}
{\bf E} = - {\bf \nabla} \phi -\frac{1}{c} \frac{\partial {\bf A}}{ \partial t}
\end{equation}
Substitution of (42) and (43) in Eq.(30) results into the wave equation for ${\bf A}$
\begin{equation}
\nabla^2 {\bf A} -\frac{1}{c^2} \frac{\partial^2 {\bf A}}{\partial t^2}=0
\end{equation}
provided
\begin{equation}
{\bf \nabla}\cdot {\bf A} +\frac{1}{c} \frac{\partial \phi}{\partial t} =0
\end{equation}
Eq.(45) is the Lorentz gauge condition. Using (45) one gets the wave equation for the scalar potential from Eq.(27)
\begin{equation}
\nabla^2 \phi -\frac{1}{c^2} \frac{\partial^2 \phi}{\partial t^2}=0
\end{equation}

It may seem that EM waves could be described by true vector waves in view of the wave equations (44) and (46) for true vector field ${\bf A}$ and true scalar field $\phi$ respectively. However, the measurable physical quantities, for example, energy density, Poynting vector, and the Lorentz force depend only on the EM fields ${\bf E}$ and ${\bf B}$. These fields are invariant under the gauge transformations of the EM potentials
\begin{equation}
{\bf A} ~\rightarrow ~{\bf A} + {\bf \nabla} \chi, ~\phi ~\rightarrow ~\phi -\frac{!}{c} \frac{\partial \chi}{\partial t}
\end{equation}
The observable physical quantities are gauge-invariant. Therefore, there is an arbitrariness in the choice of EM potentials, and there is a question of the gauge-fixing. Lack of uniqueness of EM potentials introduces additional new dimension to the nature of EM waves as vector waves. 

One may argue that the EM potentials are superfluous in the description of EM phenomena, and the issue of gauge-invariance is dispensable. 
However, in the relativistic field theory based on the action principle gauge-invariance and the Lorentz invariance are fundamental principles. The independent dynamical field variables in the action integral for EM theory are the vector potentials $A_\mu$. EM fields defined by (42) and (43) are combined to define EM field tensor
\begin{equation}
F_{\mu\nu} = \partial_\mu A_\nu - \partial_\nu A_\mu
\end{equation}
The Lagrangian density for EM fields is
\begin{equation}
\mathcal{L}_{EM} = -\frac{1}{4} F_{\mu\nu} F^{\mu \nu}
\end{equation}
The action integral
\begin{equation}
S_{EM} = \int_\Omega \mathcal{L}_{EM} d^4 x
\end{equation}
leads to the equations of motion taking the variations in the independent field variables $A_\mu$
\begin{equation}
\partial_\mu F^{\mu\nu} =0
\end{equation}
Eq.(51) corresponds to the set of field equations (27) and (30). Remaining Maxwell equations (28) and (29) follow from the definition of the EM field tensor (48)
\begin{equation}
\partial_\lambda F_{\mu \nu} + \partial_\nu F_{\lambda \mu} +\partial_\mu F_{\nu \lambda}=0
\end{equation}
Eq.(52) is known as Bianchi identity. 

The most important consequence of the action formulation is that the continuous symmetries and the invariance of the action under them lead to the conservation laws. Energy density of the EM fields and the Poynting vector emerge naturally as the components of the conserved quantities \cite{10,17}. Since the field variables $A_\mu$ and their time derivatives at each space-time are the independent variables in the action (50) where we have integration over a region $\Omega$ the conserved quantities are densities. Under infinitesimal space-time translations the invariance of (50) leads to the conserved canonical energy-momentum tensor
\begin{equation}
E^{\mu\nu}_{EM} = \frac{1}{4} \eta^{\mu\nu} F_{\alpha \beta} F^{\alpha \beta} - F^{\mu\lambda} \partial^\nu A_\lambda
\end{equation}
Invariance under infinitesimal proper homogeneous Lorentz transformations gives a conserved quantity
\begin{equation}
M^{\alpha\mu\nu}_{EM} = (E^{\alpha\nu} x^\mu -E^{\alpha\mu} x^\nu) +(F^{\alpha\nu} A^\mu -F^{\alpha\mu} A^\nu) = L^{\alpha\mu\nu} +S^{\alpha\mu\nu}
\end{equation}
One expects that the third-rank tensor $M^{\alpha\mu\nu}_{EM}$ represents the AM density of the EM fields. Comparing (54) with(16) for the scalar field $M^{\alpha\mu\nu}_{scalar}$ it is clear that the second bracketted term in (54) arises due to the vector nature of the EM field. The separation shown in (54) shows that $L^{\alpha\mu\nu}$ has a formal structure suggestive of orbital AM while $S^{\alpha\mu\nu}$ depends only on the fields $F^{\alpha\mu}$ and $A^\nu$ that has a plausible interpretation as intrinsic spin AM of the EM fields.

However, the canonical energy-momentum tensor (53) due to the explicit presence of $A^\mu$ is not gauge-invariant; it is also not symmetric. The tensor $M^{\alpha\mu\nu}_{EM}$ is also not gauge-invariant. Moreover, the separation into $L^{\alpha\mu\nu}$ and $S^{\alpha\mu\nu}$ does not preserve Lorentz invariance. Constructing a symmetric gauge-invariant energy-momentum tensor
\begin{equation}
T^{\mu\nu}_{EM} = F^{\mu\alpha} F^\nu_\alpha +\frac{1}{4} \eta^{\mu\nu} F^{\alpha\beta} F_{\alpha\beta}
\end{equation}
a natural gauge-invariant third-rank tensor can be defined
\begin{equation}
J^{\alpha\mu\nu}= T^{\alpha\nu}_{EM} x^\mu - T^{\alpha\mu}_{EM} x^\nu
\end{equation}
The tensors (54) and (56) differ by a total divergence term
\begin{equation}
M^{\alpha\mu\nu}_{EM} = J^{\alpha\mu\nu} -\partial_\lambda [F^{\lambda \alpha}(A^\mu x^\nu -A^\nu x^\mu)]
\end{equation}
The divergence term in (57) depends on the EM potentials. It is this term that was proposed to represent AM holonomy for GP in optics \cite{2}.

{\bf PHYSICAL REALITY of the EM POTENTIALS}

Intricate role of the gauge symmetry in EM field theory arises due to the 4-vector EM potential $A^\mu$ together with the belief that only gauge-invariant quantities are observable/physical. It has led to the unsolved conceptual problems in classical electrodynamics and QED. Two diametrically opposed approaches are possible: (i) only EM fields are fundamental; EM potentials are just calculational tools, and (ii) a pure vector field theory in which $A^\mu$ are fundamental fields and EM fields are derived quantities. Note that the Maxwell field equations and the charge-field interaction based on the force laws already belong to the first approach. However, formulation based on action functional and quantization of fields cannot be accompalished using EM fields as fundamental field variables. An interesting idea to construct action integral using EM fields \cite{45, 46} was given a Lorentz covariant form by Sudbery \cite{47} proposing a pseudo-vector action. Physicists seem to be uncomfortable with a vector or pseudo-vector Lagrangian density as the standard Lagrangian density is a Lorentz scalar. Note that a vector Lagrangian density does not violate relativistic invariance. A serious drawback of Sudbery's action is in the context of quantization: it is not known how to quantize this theory. 

At a classical level there are notable novel consequences of Sudbery's action. Chiral invariance or the invariance under duality rotations
\begin{equation}
{\bf E}~ \rightarrow ~{\bf E} \cos \xi +{\bf B} \sin \xi
\end{equation}
\begin{equation}
{\bf B} ~\rightarrow ~-{\bf E} \sin \xi +{\bf B} \cos \xi
\end{equation}
leads to the symmetric energy-momentum tensor $T^{\mu\nu}$ as a Noether conserved quantity. Surprisingly, the invariance under space-time translations results into a conserved third-rank tensor related with Lipkin's zilch tensor $Z^{\alpha\mu\nu}$. Lipkin \cite{48} shows that zilch density tensor for a plane monochromatic EM wave depends on the state of the polarization of the wave: for a linearly polarized wave there is no flow of zilch, however, the zilch flows at the rate proportional to the frequency of the wave and oppositely directed for LCP and RCP waves. Local duality invariant electrodynamics (LDIE) based on a generalized Sudbery's action was developed by the author \cite{49}. One of the interesting physical implications discussed in \cite{49} is that of AM holonomy; in analogy to AB effect, AM holonomy may arise due to the duality gauge potential as a topological effect \cite{50}. The remarkable thing about LDIE is that it has been formulated generalizing the Maxwell field equations as well as using the action principle \cite{49,51,52}. We draw attention to interesting papers exploring duality symmetry \cite{53,54,55,56,57,58,59,60,61}; perhaps a thorough critical study on these papers may give new insights on the nature of local duality symmetry in electrodynamics.

The second approach postulates EM potentials $A^\mu$ as fundamental physical fields. Historically, Riemann in 1861 considered Eq.(45) (Lorenz equation, now termed as Lorentz gauge condition) to interpret scalar field as density of aether, and $c{\bf A}$ as velocity \cite{62}. Later developments leading to Maxwell-Lorentz theory in which EM fields acquire overwhelming physical significane \cite{62} pushed the question of physical significance of EM potentials to the back-stage. The fundamental role of EM potentials in quantum mechanics, QED, and the Standard Model of particle physics is recognized, however, the issue of their physical reality remains unsettled.  The experimental proof of the AB effect \cite{9} strongly suggests the direct physical effect of EM potentials, however, there do exist contrary views. In a recent paper \cite{37} pure vector field for a physical photon has been discussed that gives rigorous foundation for the topological photon \cite{36}. The important question that may be posed is whether the co-existence of EM fields and EM potentials is responsible for the weird claims in the literature on the structured light beams and single photon experiments. A brief appraisal on these issues is made in the following.

{\bf STRUCTURED LIGHT BEAMS}

What is the meaning of a structured light beam? I think any light wave other than natural light could be considered as structured light. Laser light beams in existence for more than a half century present sophisticated example of structured light. Kiselev \cite{63} notes that lasers and the generation of ultra short pulses gave great impetus to the studies on wave phenomena. Though the review \cite{63} is confined to the spatio-temporal localized solutions of the scalar wave equation (3) one can discern the importance of plane waves in the context of these solutions that belong to the structured light beams. Monochromatic scalar plane wave needs just one physical parameter, that is, the propagation vector, for its charecterizaion since frequency is fixed. Plane waves propagating in different directions could serve as "bricks" for constructing complicated wave structures \cite{16}. For the vector waves additional parameter, a polarization vector $\epsilon_\mu$, is required to account for the vector nature of the wave. Superposition of plane vector waves with different $({\bf k},~\epsilon_\mu)$ would serve the basis for synthesizing the desired wave structures. 

Let us consider singular waves. Singular waves having line singularities for scalar and vector waves are defined by the vanishing amplitudes at the singularity. For the vector waves one assumes a longitudinal direction of propagation and singular lines have instantaneous zero values for the  transverse field. The terminology wave dislocation for singular scalar waves and disclination for singular vector waves originates drawing analogy with crystal dislocations and the line singularity in the pattern of crystal directions in liquid crystals respectively \cite{64}. For EM waves or light beams the vector fields $({\bf E}, {\bf B})$ may possess different pattern for disclinations. A generalized singular vector wave field originates for spatially varying polarization and different directions of propagation \cite{65}. For this kind of singular wave fields the definition of the propagation vector as phase gradient has limited validity \cite{11}.

In general, polarization ellipse, for example Eq.(24), depends on the spatial position. The electric field components can be calculated from the circular components having amplitudes $\rho_1,~\rho_2; (\rho_1>\rho_2)$, and phases $\phi_1, \phi_2$ respectively to be
\begin{equation}
E_x^\prime =Re ~ (\rho_1 +\rho_2) e^{i \epsilon -i \omega t}
\end{equation}
\begin{equation}
E_y^\prime =Re ~ i(\rho_1 -\rho_2) e^{i \epsilon -i \omega t}
\end{equation}
where $\epsilon = \frac{1}{2} (\phi_1 +\phi_2)$. Define
\begin{equation}
\frac{d \phi_1}{d{\bf r}}= {\bf k}_1 ;   ~~  \frac{d \phi_2}{d{\bf r}}= {\bf k}_2
\end{equation}
then the weighted mean is defined as 
\begin{equation}
{\bf k}_0 =\frac{ \rho_1^2}{\rho_1^2 +\rho_2^2} ~{\bf k}_1 +\frac{ \rho_2^2}{\rho_1^2 +\rho_2^2} ~{\bf k}_2
\end{equation}
On a true circular polarization $C^T$ line $\rho_2 =0$ thus ${\bf k}_0 = {\bf k}_1$. On a true linear polarization, $L^T$ line $\rho_1=\rho_2 $, therefore, ${\bf k}_0$ is a gradient of phase : $ {\bf k}_0 = \frac{1}{2} ({\bf k}_1 +{\bf k}_2) ={\bf \nabla } \epsilon$.

Scalar and vector plane waves having different propagation vectors in all directions, and polarization vectors varying at each space-time points
 could be superposed in a suitable way to construct structured light in a what could be called a bottom-up approach. Alternatively, in a top-down approach the structured light could be resolved into constituent plane waves. In both approaches, the essential irreducible physical characteristics would be frequency, propagation vector, and polarization vector. For experimental realization of the theoretically possible structured light one may impose a physical constraint: taking the guidance from \cite{34,35} the stability of the wave pattern could serve this purpose. This criterion may prove useful to exclude exotic/counter-intuitive claims on the structured light waves.

{\bf SINGLE PHOTON}

Simmons and Guttmann \cite{15} use simple arguments to show that photon localization in space is not possible; it is logical to seek photon description in momentum space. Physical properties of photon, in particular, spin and orbital AM find intuitively appealing explanation in this picture. In quantum optics a single photon of wave vector ${\bf k}$ and polarization index $s$ is represented by an excitation of the EM radiation field in the Fock space \cite{14}. Does such a single photon state represent a physical photon? The import of this question can be realized once the conceptual issues in the Fock state description of photon, most often ignored in the literature, are identified. Planck oscillator and Einstein's discrete energy light quanta serve key ideas in the quantization of EM field in both QED and quantum optics. 

In quantum optics one begins with vector wave equation and treats field variables as quantum operators. Fourier series expansion of a vector field ${\bf V}({\bf r}, t)$ 
\begin{equation}
{\bf V} = \sum_{\bf k} ( {\bf V}_{\bf k}(t) e^{i {\bf k}\cdot {\bf r}} +{\bf V}_{\bf k}^*(t) e^{-i {\bf k}\cdot {\bf r}})
\end{equation}
in the wave equation (17) leads to
\begin{equation}
\frac{\partial^2 {\bf V}_{\bf k}(t)}{\partial t^2}= - \omega_k^2 {\bf V}_{\bf k}(t)
\end{equation}
where $\omega_k = c k$. Eq.(65) resembles simple harmonic oscillator (HO) equation. It is suggestive of interpreting $V_{\bf k}(t)$ as an analogue to the position operator. A harmonic time solution of Eq.(65) 
\begin{equation}
{\bf V}_{\bf k}(t)= {\bf V}_{\bf k} e^{-i \omega_k t}
\end{equation}
substituted in the Fourier expansion (64) results into the expansion in terms of field modes $(V_{\bf k},~ V_{\bf k}^*)$. In analogy to the HO operators, the mode variables are expressed in terms of canonically conjugate variables $(q_k, p_k)$
\begin{equation}
{\bf V}_{\bf k}= N (\omega_k  q_k +i p_k) \hat{\epsilon}_{\bf k}
\end{equation}
\begin{equation}
{\bf V}_{\bf k}^*= N (\omega_k  q_k -i p_k) \hat{\epsilon}_{\bf k}
\end{equation}
Polarization unit vector $\hat{\epsilon}_{\bf k}$ takes into account the vectorial nature of the field, and $N$ is a normalization constant for the cavity modes. Recalling that the creation and annihilation operators $\hat{a}^\dagger$ and $\hat{a}$ respectively for a quantum HO are given by
\begin{equation}
\hat{a}= (2\pi \hbar \omega)^{-\frac{1}{2}} (\omega \hat{x} +i \hat{p})
\end{equation}
\begin{equation}
\hat{a}^\dagger= (2\pi \hbar \omega)^{-\frac{1}{2}} (\omega \hat{x} -i \hat{p})
\end{equation}
the cavity quantization of the vector field ${\bf V}({\bf r}, t)$ and the mode representation (67)-(68) relate $a_{\bf k}^\dagger$ and $a_{\bf k}$ with the creation and destruction of a quantum of energy $\hbar \omega_k$ and wave vector ${\bf k}$. The vector field is equivalent to an infinite sum of HOs as $k$ is a continuous variable. Introducing number operator $\hat{n}_{\bf k} =a_{\bf k}^\dagger
a_{\bf k}$ and defining a vacuum state that has no particles (excitations) of any momentum $|0> = |0,0,0,.....>$ the state with a single excitation of wave vector ${\bf k}$ and polarization index $s$ can be defined in the Fock space. Note that the vector fields ${\bf V}$, ${\bf E}$, ${\bf B}$, and ${\bf A}$ satisfy the identical wave equations (17), (31), (32), and (44) respectively, therefore, the formal construction for the quantum field operators could be carried out for any of them similar to that discussed above. Thus, to term a single particle excitation as photon becomes arbitrary and ambiguous.

Let us have a look on a different way for the EM field quantization. Similarity between the total energy of a HO with unit mass
\begin{equation}
E_{osc} = \frac{1}{2} (p^2 +\omega^2 x^2)
\end{equation}
and the energy density of the EM fields $\frac{1}{2} ({\bf E}^2 + {\bf B}^2)$ suggests following canonically conjugate field variables as quantum operators
\begin{equation}
\hat{{\bf E}} =-\sum_k p_k(t) {\bf E}_k({\bf r})
\end{equation}
\begin{equation}
\hat{{\bf B}} =\sum_k  \omega_k q_k(t) {\bf B}_k({\bf r})
\end{equation}
In this case the Fock state description utilizing both electric and magnetic field vectors gives photon interpretation as EM field quantum for single particle excitation $|0, 0, 0, ......1_{{\bf k} s}, ....>$ in the number space $\{ n\}$.

However, for the number state one has
\begin{equation}
<n|\hat{{\bf E}}|n> =0
\end{equation}
The electric field is not well defined. Introducing a coherent state $|\alpha>$ as a superposition of number states it can be shown that $<\alpha|\hat{{\bf E}}|\alpha>$ is non-zero. In quantum optics, usually one interprets electric field amplitude per photon using $
<n|\hat{{\bf E}}^2|n>$ that is calculated ignoring zero-point energy/field. Knight \cite{66} elucidates physics of the experiments on radiation trapped in a cavity. The electric field of a trapped light quantum of frequency $\omega$ in a cavity of volume $V$ has electric field amlitude in discrete units $E_0 = (\frac{\hbar \omega}{2V})^\frac{1}{2}$; its numerical value for typical experiments is of the order of millivolt per centimetre. For $n$ quanta the quantum of electric field is $E_0 ~n^\frac{1}{2}$. The experiments seem to indicate the discrete nature of photon; however, the evidence cannot be considered unambiguously direct.

It seems reasonable to assert that in quantum optics one does not need EM potentials, the word photon for a single-particle  excitation is just a convenient term, and the polarization index $s$ has no direct and physically acceptable relation with the spin AM of photon.
  
In QED, the EM field quantization proceeds with $A_\mu$  assumed as quantum field variables and defining conjugate momentum fields from the Lagrangian density $\mathcal{L}_{EM}$. To incorporate electron-photon interaction, a current-current interaction term $e\bar{\Psi} \gamma^\mu \Psi A_\mu$ is added to the Lagrangian density $\mathcal{L}_{EM}$. Here $\Psi$ is a Dirac 4-spinor field for electron. If QED is treated as a $U(1)$ gauge field theory then the interaction term originates imposing local gauge invariance on the free field Lagrangian density of electron and photon. In this approach, the principles of manifest Lorentz covariance and gauge invariance have to be respected. However, there arise insurmounable difficulties due to these principles. In addition, the problem of infinities and the renormalization method to deal with them continue to remain unsatisfactory from the aesthetic point of view of a fundamental theory in spite of precision empirical success of QED. Some of the textbooks discuss this aspect. The main point that we wish to underline is that unlike quantum optics, the role of EM potentials is indispensable in QED. Obviously, the question of gauge invariance is unavoidable in QED. 

Transverse photon as a gauge boson seems to have physical reality. However, longitudinal and time-like photons that must appear in covariant quantization are considered  'unphysical'. It is axiomatic to state that observables must be gauge-invariant. Definition of spin AM in QED remains unsettled due to the question of gauge invariance. In contrast, spin AM of laser light beams following quantum optics approach has been measured experimentally. Beginning with Beth experiment \cite{28} that measured spin AM of polarized light, the measured spin AM and orbital AM of structured light beams represent the physical properties of a large number of photons. The Fock state description allows the spin and/or orbital AM per photon inference. However, spin/orbital AM per photon cannot be taken to imply spin/orbital AM of a real physical photon. In QED, the concept of photon and its physical properties involve the intricate issues of Lorentz covariance for massless field and gauge invariance for the observables. Leader \cite{67} articulates conflict between quantum optics and QED in connection with the issue of spin and orbital AM, and seeks resolution of the conflict. However, in the light of the present arguments we suggest that the issue of conflict between quantum optics and QED is non-existent.

\section{\bf Geometry and Holonomy}

The concept of a straight line and the fifth axiom of parallel straight lines in the Euclidean geometry have been generalized to non-Euclidean geometry introducing infinitesimal parallelism and affine manifold \cite{68}. The geometry of paths \cite{69} is equivalent to the affine geometry of Weyl \cite{68}. Earlier, another concept of the Euclidean geometry, that of the distance between two points, found generalization by Gauss-Riemann to the distance between infinitesimally close points introducing a fundamental metric tensor $g_{ij}$ that is a continuous function of the coordinates, see \cite{68,70}. Riemannian geometry is a non-Euclidean geometry. In general, holonomy is intrinsic to non-Euclidean geometry. 

First, let us consider Riemannian geometry. The distance between a point $P_1(x_\mu)$ and a point $P_2(x_\mu+ dx_\mu)$ in the immediate neighbourhood of $P_1$ is postulated to be given by the line element
\begin{equation}
ds^2 =g_{\mu\nu} dx^\mu ~dx^\nu
\end{equation}
Here, $g_{\mu\nu}(x^\mu)$ is the metric tensor. The 3-index Christoffel symbols are introduced using the first-order partial derivatives of the metric tensor; the expression for one of them is given below
\begin{equation}
\Gamma^\alpha_{~\mu\nu} = \frac{1}{2} g^{\alpha \lambda} [ \frac{\partial g_{\mu\lambda}}{\partial x_\nu} + \frac{\partial g_{\nu\lambda}}{\partial x_\mu} - \frac{\partial g_{\mu\nu}}{\partial x_\alpha}]
\end{equation}
Note that the Christoffel symbol $\Gamma^\alpha_{~\mu\nu}$ is not a tensor. It is symmetric $\Gamma^\alpha_{~\mu\nu} =\Gamma^\alpha_{~\nu\mu}$.

The differential change in a vector $A_\mu$ for $x_\mu \rightarrow x_\mu +dx_\mu$ is not just the ordinary derivative
\begin{equation}
A_{\mu, \nu} = \frac{\partial A_\mu}{\partial x_\nu}
\end{equation}
but a covarianrt derivative
\begin{equation}
A_{\mu :\nu} = A_{\mu,\nu} -\Gamma^\alpha_{~\mu\nu} A_\alpha
\end{equation}
To understand the concept of holonomy let us calculate the change in a vector field $A_\alpha$ round a circuit ABCDA in the plane of $x_\mu-x_\nu$ where $A(x_\mu, x_\nu),~B(x_\mu+dx_\mu,x_\nu),~C(x_\mu+dx_\mu, x_\nu+dx_\nu),~ D(x_\mu, x_\nu+dx_\nu)$. Calculation leads to
\begin{equation}
(A_{\alpha:\nu:\mu} -A_{\alpha:\mu:\nu}) dx^\mu dx^\nu
\end{equation}
The absolute change (79) is non-zero. This change $\Delta A_\alpha$ represents a change in the direction of the vector field when it returns to the initial point A. It may be termed the direction holonomy. The holonomy is a consequence of the non-integrability of (78). The integral $\int dA_\mu$ between two points depends on the path of integration. If $d A_\mu$ is a complete differential the integral is independent of the path of integration. The condition for integrability is that $A_{\mu:\nu}$ given by (78) vanishes. 

Calculating the second covariant derivative of $A_\alpha$ in Eq.(79)  we get
\begin{equation}
A_{\alpha:\nu:\mu} -A_{\alpha:\mu:\nu} =A_\sigma R^\sigma_{~\alpha\mu\nu}
\end{equation}
Here $R^\sigma_{~\alpha\mu\nu}$ is the Riemann-Christoffel curvature tensor in which the second-order partial derivative of the metric tensor appears. The integrability condition becomes the vanishing of the curvature tensor
\begin{equation}
R^\sigma_{~\alpha\mu\nu} =0
\end{equation}
The metric geometry with the condition (81) reduces to the Euclidean geometry: uniform direction can be given at all points in the plane.

The significance of the Christoffel symbol becomes manifest in the equations of a geodesic. A geodesic is a path for which $\int ds$ is stationary
\begin{equation}
\int \delta(ds) =0
\end{equation}
For the line element (75) one finally arrives at the geodesic equation
\begin{equation}
\frac{d^2 x^\alpha}{ds^2} + \Gamma^\alpha_{~\mu\nu} dx^\mu dx^\nu =0
\end{equation}

Now, let us consider the geometry of paths. A straight line in the Euclidean geometry is described by a differential equation
\begin{equation}
\frac{d^2 x^\mu}{ds^2} = 0
\end{equation}
where $s$ is a parameter defining the motion of a point $x^\mu(s)$. Generalization of Eq.(84) to
\begin{equation}
\frac{d^2 x^\mu}{ds^2} + ^a \Gamma^\mu_{~\nu \sigma}~ \frac{dx^\nu}{ds}~ \frac{dx^\sigma}{ds} =0
\end{equation}
\begin{equation}
^a\Gamma^\mu_{~\nu \sigma}=^a\Gamma^\mu_{~\sigma \nu}
\end{equation}
represents a general theory of the geometry of paths. The coefficients $^a\Gamma^\mu_{~\nu \sigma}$ are analytic functions of the coordinates $x^\mu$. A coordinate transformation $x^\mu ~\rightarrow ~f^\mu(y^\nu)$ in Eq.(84) results into the form of Eq.(85); comparing the coefficients in the second term of the transformed Eq.(84) and Eq.(85) we identify
\begin{equation}
^a\Gamma^\mu_{~\nu \sigma} =\frac{\partial y^\mu}{\partial x^\alpha}~ \frac{\partial^2 x^\alpha}{\partial y^\nu \partial y^\sigma}
\end{equation}
Conversely it may be asked whether Eq.(85) can be transformed to Eq.(84). The condition for this to happen turns out to be the vanishing of the curvature tensor
\begin{equation}
B^\alpha_{~\mu\nu\sigma} =\frac{\partial ^a\Gamma^\alpha_{~\mu \nu}}{\partial x^\sigma} -\frac{\partial ^a\Gamma^\alpha_{~\mu \sigma}}{\partial x^\nu}+{^a\Gamma^\beta_{~\mu \nu}} ^a\Gamma^\alpha_{~\beta \sigma}-{^a\Gamma^\beta_{~\mu \sigma}} ^a\Gamma^\alpha_{~\beta \nu}
\end{equation}

An affine manifold is defined  based on the infinitesimal parallel displacement of a vecor, i. e. Levi Civita -Weyl parallelism \cite{68}. An affine relation in this manifold is defined by the transformation of a vector at a point to its neighbourhood by infinitesimal parallel displacement of the vector. Let $A^\mu$ be a vector at the point $x^\mu$, and $A^\mu+dA^\mu$ is the vector after parallel displacement to the point $x^\mu +dx^\mu$. In a linear affine relationship the change $dA^\mu$ depends linearly on $A^\mu$ with the coefficients $^a\Gamma^\mu_{~\nu\sigma} dx^\sigma$. The affine connection is an analytic function of the coordinates, and every point of the manifold is affinely related to its neighbourhood. A coordinate transformation can make the affine connection to vanish at a single point, however, it cannot be made to vanish for all points simultaneously. The condition for the parallel displacement of a vector $A^\mu$ is that $\frac{\partial A^\mu}{\partial x^\nu} + ^a\Gamma^\mu_{~\nu\lambda} A^\lambda =0$.

If a point $x^\mu(s)$ of the manifold has a vector $A^\mu$ associated with it then this vector is stationary at $s$ if 
\begin{equation}
\frac{dA^\mu}{ds} + ^a\Gamma^\mu_{~\nu\sigma} A^\nu \frac{dx^\sigma}{ds} =0
\end{equation}
Eq.(89) defines the parallel displacement along the trajectory of the point $x^\mu(s)$ as the parameter $s$ varies. It is clear that the geometry of paths involves the affine connection $^a\Gamma^\mu_{~\nu\sigma}$. Remarkably, the definition of affine connection (87) and the curvature tensor (88) do not require the metric tensor. However, the condition for the transformation to the Euclidean geometry is identical in both affine geometry of Weyl and the metric geometry of Riemann: the vanishing of the curvature tensor. If the geometry of paths or affine manifold has a metric structure, the affine connection and Christoffel symbols are equivalent. 

The non-integrability noted by Weyl \cite{68} for Eq.(79) can be re-written introducing the surface element $dS^{\mu\nu}$ for the displacements $dx^\mu$ and $dx^\nu$
\begin{equation}
\delta A_\alpha = \frac{1}{2} (A_{\alpha:\mu:\nu} -A_{\alpha:\nu:\mu}) ~dS^{\mu\nu}
\end{equation}
Using Eq.(80) the expression on the RHS of Eq. (90) can be re-written as $\delta A_\alpha = \frac{1}{2} R_{\alpha\mu\nu\sigma} A^\sigma dS^{\mu\nu}$. The change in the length of the vector $A^\alpha A_\alpha$ becomes
\begin{equation}
A^\alpha\delta A_\alpha = \frac{1}{2} R_{\alpha\mu\nu\sigma} A^\alpha A^\sigma dS^{\mu\nu}
\end{equation}
Since $R_{\alpha\mu\nu\sigma}$ is anti-symmetrical in the indices $(\alpha, \sigma)$ the change (91) vanishes. Thus, the non-integrabilty (90) results into the change in the direction of the vector parallel transported round a circuit keeping the length of the vector unchanged.  

Weyl \cite{68} introduced non-integrability of the length of a vector in a generalized geometry, let us call it  Weyl geometry, in which there is a quadratic ground form for the metric  $g_{\mu\nu} dx^\mu dx^\nu$, and a linear form $k=k_\mu dx^\mu$. In Weyl geometry, there is a wider group of symmetry transformations: general coordinates transformations and gauge transformations. In this geometry, both direction and length holonomy occur. 

The holonomy is an important characteristic of local differential geometry beginning with the Gauss-Riemann geometry to the Levi-Civita parallelism and Weyl's affine manifold. One could view a family of affine spaces as the fibers and an affine connection depending on the curves in the base space. Weyl geometry has additional non-compact group of homothetic transformations: the line element under the gauge transformations $ds^2 ~\rightarrow ~ \lambda ds^2$ and $k~\rightarrow ~ k+ d \log \lambda$. In the language of fiber bundles, the later version of Weyl gauge theory that has found application in the Standard Model of particle physics, is a circle or a $U(1)$-bundle; instead of  scale or gauge transformation here one deals with the phase transformations.

The geometry at large or global aspects of the differential geometry have important consequences for the understanding of the mathematically natural geometries. One of the earliest examples is that on the Gauss-Bonnet formula: consider a closed orientable surface, then the integral of the Gaussian curvature is equal to $2\pi \chi$, where $\chi$ is the Euler characteristic of the surface. 

An insightful derivation of the Gauss-Bonnet formula by Chern \cite{71} may be useful for physical applications. A linear differential form or 1-form is interpreted as the connection form in the fiber space $B$. Let us consider an orientable 2D surface $M$. The 3D fiber space B is a space of all unit tangent vectors on the surface $M$. What is the connection 1-form in $B$? To specify orientation, a unit tangent vector $\xi=(\xi^1, \xi^2)$ and an orthogonal unit tangent vector $\eta =(\eta^1, \eta^2)$ form a positive orientation. The connection 1-form is defined to be
\begin{equation}
\phi = g_{ij}~D\xi^i \eta^j 
\end{equation}
where the covariant differential is
\begin{equation}
D\xi^i =d\xi^i + \Gamma^i_{~jk}~ \xi^j du^k 
\end{equation}
The surface $M$ has the line element in the isothermal coordinates
\begin{equation}
ds^2 = e^{2 \lambda(u,v)} (du^2 +dv^2)
\end{equation}
Here the index $i=1, 2$ and the notation $(u^1=u, ~u^2 =v )$ is convenient for calculations. Eq.(94) is re-written as $ds^2 = g_{ij} du^i du^j$. The following values of the Christoffel symbol  are obtained
\begin{equation}
\Gamma^1_{~11} =\Gamma^2_{~12} =-\Gamma^1_{~22} =\lambda_u
\end{equation}
\begin{equation}
\Gamma^1_{~12} =\Gamma^2_{~22} =-\Gamma^2_{~11} =\lambda_v
\end{equation}
Here, the partial derivatives of $\lambda$ with respect to $u^1, u^2$ are denoted by $\lambda_u, \lambda_v$ respectively. 
The expression for $D\xi^i$ (93) is evaluated using (95) and (96)
\begin{equation}
D\xi^1 = d\xi^1 +(\lambda_u du +\lambda_v dv)\xi^1 + (\lambda_v du -\lambda_u du)\xi^2
\end{equation}
\begin{equation}
D\xi^2 = d\xi^2 +(\lambda_u dv -\lambda_v du)\xi^1 + (\lambda_u du +\lambda_v dv)\xi^2
\end{equation}

Choosing the set of the orthogonal unit vectors 
\begin{equation}
\xi^1 =e^{-\lambda} \cos \theta, ~  \xi^2 =e^{-\lambda} \sin \theta ;~ \eta^1 =e^{-\lambda} \sin \theta, ~ \eta^2 =e^{-\lambda} \cos \theta
\end{equation}
the connection 1-form (92) is obtained to be
\begin{equation}
\phi =d\theta -\lambda_v du +\lambda_u dv
\end{equation}
The exterior derivative of the differential form (100) is
\begin{equation}
d\phi = - K ~dA
\end{equation}
where the area element
\begin{equation}
dA = e^{2\lambda} du dv
\end{equation}
and the Gaussian curvature of the surface $M$ is
\begin{equation}
K = - e^{2\lambda} (\lambda_{uu} +\lambda_{vv})
\end{equation}           
Chern \cite{71} remarks that Eq.(101) is "perhaps the most important formula in two-dimensional local Riemmanian geometry".

The proof of the Gauss-Bonnet theorem is given in \cite{71}, we only state the theorem: Let $D$ be a compact oriented domain in $M$ bounded by a sectionally smooth curve C. Then 
\begin{equation}
\int_C k_g~ ds +\int_D K~dA+\sum_i (\pi -\alpha_i)=2\pi \chi
\end{equation}
where $k_g$ is the geodesic curvature of $C$, $(\pi -\alpha_i)$ are the exterior angles of the vertices of $C$ and $\chi$ is the Euler characteristic of $D$.

To get an idea of the indices of vector fields with isolated singularities we list six cases of the singularities: a source or a maximum, a sink or minimum, a center, a simple saddle point, a monkey saddle, and a dipole. The respective indices are 1, 1, 1, -1, -2, 2. To define geodesic curvature $k_g$, let us consider a smooth unit vector field $\xi(s)$ along the curve $C$; $s$ is arc length. If  $\xi(s)$ is everywhere tangent to the curve, then $\phi =k_g ds$. If the geodesic curvature is zero, $C$ is a geodesic.

\section{\bf Holonomy: Geometry to Physics}

Role of mathematics in physics was discussed in Newton’s Principia that represents “the mathematical principles of philosophy”.  Poincare's insights on mathematical and scientific truths, and the commonly accepted view that the mathematics is the natural language of physics may have to be critically examined with a simultaneous study of differing views held by prominent physicists reported in \cite{32}. In the monograph "Time-Transcendence-Truth" \cite{72} a radically new idea is put forward: "The mental creations in mathematics find nearest expression in the tangible world of physics." This idea forms the basis for the hypothesis that physics is the natural language of mathematics \cite{73}. 

The concept of transcendence used in my book \cite{72} needs to be placed in the right perspective comparing it with that used in Weyl's book \cite{68}. According to Weyl, the immediate observable world appears as intentional reality in the stream of consciousness. It has transcendental reality, i. e. the phenomenal existence. On the other hand, the intentional object that appears as the primary act of perception acquires a kind of immanence or absoluteness on the second act of reflection. Weyl believes that time is 'the primitive form of the stream of consciousness'. In my view, transcendence is beyond the consciousness. If we assume the idea of sub-conscious mind to be the subtler state of mind beyond the consciousness then transcendence originates in the sub-conscious mind. Transcendental reality presents itself to the conscious mind as abstract mathematical truth. The mathematical truth so perceived finds expression in the form of physical reality through sense-experiences. Thus physics is the language of mathematics, and by its very nature represents approximate and incomplete truth. The flow of time is the fundamental transcendental reality that in the form of natural numbers emerges as the mathematical truth. Geometry and space are the creations of the action of time: time is action. 
 
The mathematical objects, for example, circle and sphere could only be realized physically in approximate and gross forms. In 3D Euclidean space, a curve is a 1D point-manifold; a point $x_i(u), ~i=1,2,3$ traces a curve if the coordinates vary as the continuous functions of a parameter $u$. Terming the parameter $u$ as 'time' the motion is defined by $\frac{dx_i}{du}=v^i$. Weyl remarks that ,'In default of a better expression we shall apply this name in a purely mathematical sense'. In physics, motion for a point mass is defined as the rate of the change of displacement; ratio of infinitesimal displacement and time interval tending to the limit zero. Obviously, physical motion is an approximation to the mathematical motion. Newton's second law of motion
\begin{equation}
\frac{d{\bf p}}{dt} = {\bf F}
\end{equation}
needs to be treated as a mathematical law, and we need the corresponding law in the language of physics. To gain new insights let us consider a force-free motion setting ${\bf F} =0$ in Eq.(105)
\begin{equation}
\frac{d{\bf p}}{dt} =constant
\end{equation} 
In the standard interpretation different values of the constant in Eq.(106) define kinematically equivalent motions. While mathematically it is true that the value of the arbitrary constant has no significance, in a physical motion with different uniform momenta there must exist a physical mechanism to cause differing momenta.  We argue that the physical motion is statistically averaged random motion: the constants in Eq.(106) are determined by the averaged density of the scattering points. Thus, physically no two inertial frames are equivalent. In a general case, the geometry of paths represented by Eq.(85) is proposed to have the physical realization in terms of an equation of motion for a stochastic process. The affine connections in the geometry of paths have to be interpreted in a statistical manifold. Motivated by the stochastic interpretation of quantum mechanics Nelson \cite{74} has discussed the theory of Brownian motion for both Einstein-Smoluchowski and Ornstein-Uhlenbeck processes. Luis de la Pena has pursued the stochastic quantum theory for past many decades \cite{75}. Here, the new idea is that irrespective of classical or quantum descriptions the statistical affine manifold determines the dynamics for physical systems.

A 2D point-manifold defines a surface in terms of two parameters $x_i(u_1,u_2), ~ i=1,2,3$.  The metric tensor $g_{ij}$ in Eq.(75) in a metric space will represent a statistical metric space in physical realization. It may be of interest to consider Menger's idea in this context \cite{76}. Statistical affine manifold and statistical metric space rather than mathematical geometric spaces describe the physical phenomena: it becomes necessary to postulate physical space consisting of scattering points distributed in the whole space; it is a kind of variant of aether \cite{62}. 

There may be another physical process originating from the non-trivial topological objects. A force-free motion results if the potential function in the assumed force ${\bf F}={\bf \nabla}V$ is constant. Arbitrary constant values of $V$ describe the equivalent motions. However, a non-trivial nature of the force that vanishes everywhere except on a 1D line signals a toplogical defect. Instead of the Newtonian equation of motion it is better to use the Hamilton-Jacobi equation
\begin{equation}
H(q_i, ~\frac{\partial S}{\partial q_i}) =E
\end{equation}
\begin{equation}
H(q_i, ~\frac{\partial W}{\partial q_i}) +\frac{\partial W}{\partial t} =0
\end{equation}
The notation $W$ for the Hamilton's principal function and $S$ for the characteristic function is that used in \cite{77}. The canonical momenta are $p_i=\frac{\partial S}{\partial q_i}$, the line integral for a closed path vanishes
\begin{equation}
\oint p_i~dq_i =0
\end{equation}
In the old quantum theory, Bohr-Sommerfeld quantization is based on the integrals $\oint p_r~dr$ and $\oint p_\theta ~d\theta$. Born and Wolf \cite{11} on page 736 underline the fact that Eq.(109) holds only if the closed loop is in the simply connected region. For multi-valued momenta the integral may have a non-zero value equal to an integral multiple of a constant. An insightful discussion on Hilbert's treatment of the Hamilton-Jacobi theory shows that the Stokes theorem breaks down if there is a non-trivial topology: though ${\bf \nabla} \times {\bf A}=0$ the integral $\oint {\bf A}\cdot {\bf dl}$ is non-zero. 

Thus both geometric and topological holonomies have physical manifestations. A remarkable result pointed out by Thomas \cite{21} somehow remains unnoticed \cite{22}. In the relativistic kinematics of an electron it is found that the successive Lorentz transformations without rotation result into a Lorentz transformation and a rotation \cite{21}. Most of the textbooks explain the Lorentz group $SO(3,1)$ and Thomas precession. If two inertial frames moving with a relative speed $v$ are considered, pure boost Lorentz transformation with the boost generators ${\bf K}$ can be defined. The boost generators do not form a closed Lie algebra. Note that the rotation generators ${\bf J}$ form a closed Lie algebra for the rotation group. The commutators between ${\bf J}$ and ${\bf K}$ are utilized to prove the result of Thomas. However, Thomas in Section 3 of his paper following Eq.(35) remarks: " This means, inter alia, that if an electron were to move in a closed path, and its axis of rotation at each moment did not change its direction in the system of coordinates in which the electron was instantaneously at rest, yet after a cycle the direction would be different." It is one of the earliest examples of parallel transport holonomy; its link with spin has been suggested in \cite{22, 23}.

Note that the geometric phases in optics do not require Berry phase paradigm of nonrelativistic quantum theory \cite{2}. Cisowski et al \cite{25} claim that "a plethora of geometric phases have been witnessed in optics", and "these are rarely linked to fiber bundles, causing key concepts such as connection and curvature to be surrounded by an aura of mathematical mystery". GP in optics at a basic level has only two kinds of processes, namely, the propagation vector changes and polarization vector changes, therefore, the plethora of geometric phases can be explained based on these two irreducible basic GPs. Unfortunately, the discussion on fiber bundles in \cite{25} depends heavily on the Berry phase, and the link with GP in optics lacks clarity making confusing connections with the mathematics of fiber bundles. In the previous section a simple example of a 2-sphere and the use of fiber space given by Chern has been presented. For a better discussion on fiber bundles we refer to the appendix "Monopoles, holonomy and fiber bundles" on page 117 in \cite{1}, and the reviews \cite{8, 24}. In fact, mathematical aspects need to be expressed in the language of physics. For this purpose we first give a brief discussion on the AB phase and AB effect that exemplify geometric and topological holonomies respectively.

{\bf Aharonov-Bohm Phase and Effect}

In the original article \cite{19} Aharonov and Bohm give a careful and thorough discussion on the problem of electron motion, and electron and EM field interaction in the presence of EM potentials for the EM field-free regions including the scattering problem. One of the problems discussed by them is that of the electron motion in a multiply-connected region caused by a magnetic flux confined in a cylindrical tube idealized as a magnetic flux line. An electron beam splitted into two parts is made to encircle the magnetic flux line in opposite directions and made to re-combine. Representing the electron beam by a Schr$\ddot o$dinger wavefunction the recombined beam shows the quantum interference depending on the AB phase difference. The Hamiltonian for this problem is
\begin{equation}
H_{AB} = \frac{({\bf p} -\frac{e}{c} {\bf A} )^2}{2m}
\end{equation}
and the wavefunctions for the two parts along the two paths $(i=1,2)$ are given by
\begin{equation}
\Psi_i =\Psi_i^0 ~ e^{\frac{-iS_i}{\hbar}};~ S_i = \frac{e}{c} \int_i {\bf A} \cdot {\bf dl}
\end{equation}
These wavefunctions are defined in simply-connected regions.

Two important points are made here:

 [P1] In the Hamiltonian (110) it is assumed that the vector potential is determined by the magnetic flux line; the flux $\Phi_0 = \oint {\bf A} \cdot {\bf dl}$ is constant but ${\bf A}$ is non-zero in the region outside the flux line. The AB effect is defined for a closed loop encircling the flux line. Therefore, the interference between the two beams depends on the phase difference
\begin{equation}
\frac{(S_1 -S_2)}{\hbar}= \frac{e}{\hbar c} \Phi_0
\end{equation}
On the other hand, the AB phase can be defined along a path between any two points 
\begin{equation}
S_{AB} = \frac{e}{c} \int_{P_1}^{P_2} {\bf A} \cdot {\bf dl}
\end{equation}
The phase due to the path difference between two points has two parts: the phase acquired in the absence of the solenoid and the phase $S_{AB}$ for the magnetic field-free region but depending on the pure gauge potential. Changing the magnetic flux will change this additional phase. In sharp contrast to this trivial geometric AB phase, the AB effect has topological origin since the electron encircles the magnetic flux-line.  

[P2] The form of the Hamiltonian (110) is unchanged for an arbitrary magnetic field such that the vector potential is determined by ${\bf B} = {\bf \nabla} \times {\bf A}$. One can use the path-dependent phase factor 
\begin{equation}
\Delta_{P_1P_2} =e^{ \frac{ie}{\hbar c} \int_{P_1}^{P_2} {\bf A} \cdot {\bf dl}}
\end{equation}
to describe the electron motion in a magnetic field. The phase factor $\Delta_{P_1P_2}$ is termed Fock-London-Weyl phase factor in \cite{9}.
Note that the the vector potential in $\Delta_{P_1P_2}$ is determined by a non-zero magnetic field, whereas for the AB phase factor the vector potential is determined for zero magnetic field region. As an example, in the Landau problem one has a uniform magnetic field along z-axis, and the vector potential in a symmetric gauge is given by
\begin{equation}
{\bf A} = \frac{B}{2} (-y \hat{x} + x \hat{y} +0 \hat{z})
\end{equation}
The role of gauge-invariance in the Landau problem has interesting dimensions, and one can choose vector potential in an alternative gauge, for example, the Landau gauge.  Wakamatsu has devoted great efforts to investigate this problem; we refer to a recent comprehensive paper \cite{78}. Symmetry group for the Landau problem is examined in \cite{79,80}.

Physical interpretation of the AB effect continues to remain unsettled \cite{9} in spite of the experimental demonstration of this effect in the pioneering work of Tonomura and his group; perhaps conclusively in 1986 \cite{81}. Physical reality of EM potentials, nonlocal interaction, and classical versus quantum nature of the effect are some of the issues that find differing view-points. We have asked a different question: what is the physical mechanism for the AB effect?  The topological AB effect has been explained in terms of the modular angular momentum exchange \cite{9}. To fully appreciate the significance of this idea we have to understand how the direction holonomy is related with the geometric/topological phase, and whether the physical system is a single electron or an electron beam. Aharonov-Bohm use the term electron beams in \cite{21}, while Tonomura et al \cite{81} and many physicists use the word electron wave for the physical system that shows the interference pattern. The shift in the whole interference pattern as the enclosed magnetic flux is changed is a measurable quantity \cite{21,81}. The interpretation of the Schr$\ddot{o}$dinger wavefunction that it describes a single electron and the intrinsic complex nature of the wavefunction implies that the AB effect is a quantum phenomenon; note that the interference of electron wave in a double slit experiment invoking wave-particle duality and Heisenberg indeterminacy a la Copenhagen interpretation \cite{41} are used here. 

Departing from this picture we argue that for a single electron motion in the field-free region the magnitude of the momentum vector of the electron does not change, the motion is along the arc of a circle, and the change in its direction can be represented as a phase in a complex plane. Thus, the direction holonomy corresponds to the geometric phase. However, mapping the arc to a straight line can be physically implemented compensating the effect of a pure gauge potential by making a gauge transformation. Physical mechanism for the geometric AB phase is suggested in terms of the statistically averaged random motion that depends on the pure gauge potential. The basis for our suggestions is drawn from heuristic arguments on the unform circular motion \cite{80} and F$\ddot{u}$rth analogy on Schr$\ddot{o}$dinger wave equation and diffusion equation \cite{41,74}.

Briefly stated, it may be recalled from elementary physics course that in a uniform circular motion the magnitude ofthe  tangential velocity is constant and the motion of the circulating point can be equivalently represented in terms of the motion of a 2D isotropic harmonic oscillator in the cartesian coordinate system. The displacement $x(t),~y(t)$ can be given a complex phasor representation $z(t) =x(t)+i y(t)=a e^{i(\omega t+ \phi_0)}$. The phasor representation maps the change in the direction in terms of the phase. If we consider the electron motion that changes the direction of uniform momentum randomly then an ensemble of single electrons with different momenta can be envisaged. Just like the phasor representation of the 2D oscillator motion a plane wave complex representation for the random motion of the electron can be given. It may, in fact, also provide further insight into the stochastic approach to quantum mechanics \cite{41, 74, 75}. Now, the geometric AB phase will be related with the difference between the actual distance traversed by the electron and the averaged distance of the scattered motion in the presence of the pure gauge potential. Changing the magnetic flux will change the density of the scattering points, and hence the averaged path-distance. An experiment performed with a beam of electrons in the double-slit experiment will show the statistical distribution of the electrons, and this whole statistical distribution pattern will shift when the enclosed magnetic flux is changed.  

Though we need a method of calculation to get quantitative results that can be compared with the measurements, the qualitative discussion does give physical understanding recalling Tonomura et al double-slit experiment for single electrons \cite{82}. This experiment was performed on the single electrons building up the interference pattern over a period of about 20 minutes. The remarks made by the authors are worth reproducing: "These results unambiguously demonstrate wave-particle duality of electrons. On the one hand, a single electron passes through two slits as a wave and forms a probability interference pattern; .......At the detector, on the other hand, an electron is observed as a localized particle. We must conclude that a certain position on the screen is selected onto which the electron wavefunction collapses. The position cannot be predicted, but occurs in the probabilistic way dictated by the probalility amplitude."  The intermediate statements that electron wave passes through two slits, and that wavefunction collapses at the point on the screen where electron is detected  are suppositions/assumptions, and cannot be proved; these follow from the acceptance of the Copenhagen interpretation. The total number of electrons to observe interference fringes is about 70000, therefore, the statistical distribution of the electrons passing through one of the slits randomly can give look-a-like wave-interference pattern. Of course, one needs a theory for such a probability distribution based on stochastic process that must account for the observations, and also agree with the quantum probability. One of the intricate questions is that of the nature of the scattering points. Non-quantized zero-point radiation field (ZPF) has been explored in \cite{75}, however, this issue remains unsettled. 

In the light of the proposition that AB phase and the AB effect arise for a single electron motion, the origin of the topological AB effect also has to be re-visited. As noted earlier, Hilbert pointed out that the Stokes theorem in a nontrivial topology could break down. Bohr-Sommerfeld quantization also points to topological quantization. Two points are involved: instead of a differential law one has an integral law, and a nontrivial multiply connected space is envisaged.  Instead of the Newtonian equation of motion or the Hamilton-Jacobi differential equation, let us consider the following integral
\begin{equation}
\oint ({\bf p} -\frac{e}{c}{\bf A})\cdot {\bf dl}
\end{equation}
If the topological quantization for the momentum integral is postulated as a multiple of $\hbar$, then the AB effect necessarily admits angular momentum exchange in view of the above integral. It has been argued in \cite{9} that pure gauge potential carries angular momentum. 

Is electron spin a topological object? Note that a remarkable electron motion holonomy has been pointed out in \cite{21}, and we have suggested that spin of the electron may have topological origin \cite{22}. In a recent paper \cite{23}, the azimuthal vector potential for a $Z_2$ vortex \cite{83} of pure geometric origin has been related with spin of $\frac{\hbar}{2}$. At a deeper level, the Pauli algebra of the group SU(2) is suggested to possess hidden topological structure based on the 2D area defect \cite{23}. Thus, physical holonomy seems to have significance in unraveling the fundamental questions in physics.

{\bf Phase Holonomy: New Results }

New physical insights into the nature of waves and light have been obtained in the preceding sections re-visiting  fundamental issues related with the phase singularities in scalar waves, and EM potentials, EM fields, and polarization singularities in optics. Utilizing these insights remarkable new results are obtained on the geometric phases and angular momentum holonomy; we present them in this section. 

{\bf RA:~}   How does one define and determine the propagation of a wave? A continuous, infinite, monochromatic plane wave for scalar disturbance or a scalar field has a mathematically well-defined propagation vector ${\bf k}$
\begin{equation}
\Phi = \Phi_0 e^{i({\bf k} \cdot {\bf r} -\omega t)}
\end{equation}
However, experimentally one needs a finite wave for determining the physical properties of the wave: velocity, energy, and momentum or propagation vector. Sommerfeld and Brillouin in 1910-1914 carried out detailed investigation on the wave propagation \cite{84} discussing the concepts of energy transport velocity, signal velocity, and other kinds of velocities. A notable point concerns the definition of a signal: it is arbitrary as the detection of a signal depends on the efficiency of a detector.  In the complex representation of a plane wave (117) it is obvious that the propagation vector is a phase gradient, and the wave velocity is phase velocity defined by setting the differential of phase equal to zero.  

The spatial part of phase in Eq.(117) can be re-written as  ${\bf k} \cdot {\bf r}=\frac{1}{\hbar} ({\bf p} \cdot {\bf r})$. We have explained that the Bohr-Sommerfeld quantization in the old quantum theory is based on the generalization of Eq.(109) for multiply-connected regions such that the integral has non-zero values. The structure of the phase in Eq.(117) is generalized to introduce an additional integral phase  
\begin{equation}
\gamma_{ph} = \int {\bf k} \cdot {\bf dr}=\frac{1}{\hbar} \int ({\bf p} \cdot {\bf dr})
\end{equation}
Eq.(114) for the phase factor in connection with the AB phase is of this kind of integral phase. If there exists a transformation of the propagation vector in the ${\bf k}$-space that gives non-zero value to the phase (118) along a geometric path in this space then it defines a geometric phase for a scalar wave. The important point is that the absolute phase of a wave is not observed as the interference of waves measures only the phase difference. For example, the initial constant phase in Eq.(117) is arbitrary and can be set to zero. However, comparing the phases of two waves with $\gamma_{ph} =0$ and non-zero $\gamma_{ph}$ it would be  possible to determine the GP of a scalar wave. The equivalence with the momentum integral in Eq.(118) shows that GP is intrinsically related with AM: the phase holonomy is equivalent to AM holonomy.  GP may possess topological attributes if the closed loop integral for the phase is non-zero
\begin{equation}
\gamma_{top} =\oint {\bf k} \cdot {\bf dr}=\frac{1}{\hbar} \oint ({\bf p} \cdot {\bf dr})
\end{equation}
It may be reminded that the scalar field has orbital AM density defined by $J^{0ij}$ in Eq.(16). A scalar wave could be given a generalized amplitude and phase structure utilizing the basic concept of propagation vector and frequency of a plane monochromatic scalar wave \cite{11, 16}. In the literature, specially in connection with the phase singularities and dislocations the main focus is on the nature of phase gradients \cite{11, 64}. 

An important class of essentially scalar waves is that of the linearly polarized laser modes \cite{29, 85}.  Let us consider the complex scalar field for LG laser light beam
\begin{equation}
E = u(r, \phi, z) e^{i(k z -\omega t)}
\end{equation}
where
\begin{equation}
u(r, \phi, z)= C A(r, z) e^{-\frac{r^2}{W^2(z)}} ~e^{i l \phi +i \Phi(r,z)}
\end{equation}
Here $C$ is a constant, $W(z)$ is the radius of the cylindrical beam, $A(r,z)$ is the amplitude function and $\Phi(r,z)$ is a phase function. LG modes are the solutions of the paraxial wave equation \cite{29, 85}. It is assumed that the vector potential is 
\begin{equation}
{\bf A} = E \hat{x}
\end{equation}
Using the Lorentz gauge condition, the calculated electric and magnetic field vectors in the paraxial approximation are utilized for deriving the expressions of the time-averaged Poynting vector and orbital AM density of the LG laser modes \cite{29}. The scalar field $u(x,y,z)$ is a solution of the paraxial wave equation in the cartesian coordinate system giving rise to HG-modes, and $u(r,\phi,z)$ is a solution of the same paraxial wave equation in the cylindrical coordinate system that leads to the LG-modes. The scalar wave (120) has line singularities called dislocations where the amplitude is zero and the phase is indeterminate. The spatial gradient of the phase is
\begin{equation}
{\bf \nabla} (k z + l \phi + \Phi) = k \hat{z} +\frac{l}{r} \hat{\phi} +{\bf \nabla } \Phi
\end{equation}
In the phase gradient equation (123) $l$ defines the topological charge of the phase vortex, and the ratio of the flux of orbital AM to that of energy is $l/\omega$. Drawing analogy to quantum theory and using the word photon allegorically it is suggested \cite{29} that LG-modes carry orbital AM of $l \hbar$ per photon. We emphasize two points: the linearly polarized laser modes (both HG and LG) represent scalar waves, and the orbital AM of the HG-modes is zero in contrast to the nonvanising  orbital AM characterized by the index $l$ for the LG-modes calculated using the classical EM theory. 

Now, we re-examine the GP associated with the mode transformations \cite{3, 39}. In analogy to the polarization state space of the light described by the Poincare sphere, the mode-state space is constructed and a new GP for the transverse modes of the orbital AM carrying beams is proposed \cite{3} and measured in an interesting experiment \cite{39}. Authors mention the role of AM transfer in this kind of GP in the mode-state space.  Physical arguments were given to show that the GP arising from the mode transformations is not new, but that based on the transformations in the momentum state space or ${\bf k}$-space \cite{6}. It seems the analogy to RVCW or spin redirection phase \cite{6} is the source of confusion; moreover, the import of the AM holonomy conjecture is not fully realized. 

The proposed integral phases $\gamma_{ph}$ and $\gamma_{top}$ resolve this problem unambiguously. In the spin redirection phase the parallel transport of the wave vector ${\bf k}$ on the surface of the sphere $S^2$ preserving the helicity ${\epsilon} \cdot \hat{\bf k}$ results into the rotation of the polarization vector after the completion of a cycle. The magnitude of the rotation is given by the solid angle subtended by the cycle, and the sign depends on the initial polarization. Left and right circular polarizations acquire equal but opposite geometric phases. On the other hand, the geometry of the spherical surface in the mode transformations is defined by
\begin{equation}
k_x^2 + k_y^2 +k_z^2 = constant
\end{equation}
such that the propagation direction is fixed, i. e. the z-axis, and the light beam remains in a fixed polarization state. For example, in the Galvez et al experiment \cite{39} the Gaussian light beam has vertical linear polarization. The time-averaged orbital AM for the Gaussian beams is also along z-axis \cite{29}, therefore, making the mode transformations for the cycle $[1\hbar \rightarrow 0 \hbar \rightarrow -1 \hbar \rightarrow 0\hbar \rightarrow 1\hbar]$ of orbital AM states or any other cycle such as $[\ -1 \hbar \rightarrow 0\hbar \rightarrow -1\hbar]$ involves the transformations in the transverse momenta $\hbar k_x, \hbar k_y$ on the spherical surface in the wave vector space. Since there are two degrees of freedom here on the sphere with fixed $k_z$ the GP is equal to half of the solid angle enclosed by the cycle. The orbital AM exchange for the mode conversion discussed in \cite{3, 39} is not AM holonomy; the AM holonomy is defined by Eq.(119).

An important experiment on the optical vortex beam \cite{86} is suggested as the direct evidence for the AM holonomy. The experiment shows that for a noncanonical vortex the topological charge undergoes inversion in the free space propagation. There are no intermediate optical elements to affect AM changes, a linearly polarized light beam is splitted into two beams, and one of them is nested with single charge screw dislocation. Noncanonical vortex dynamics is generated using a cylindrical lens. The beam transforms to an edge-line dislocation, and evolves into a vortex with opposite topological charge. Authors \cite{86} make remarks on the the question of orbital AM: the beam incident on the lens has the charge $l=+1$, the beam charge is inverted to $l=-1$ but the orbital AM is unchanged. This seemingly counter-intuitive result is explained in terms of the redistribution of orbital AM of the LG modes carrying different charges  keeping the total value constant. The alternative mechanism proposed here is that of the GP defined by Eq.(118) and the associated AM holonomy: the re-distribution of orbital AM is basically a transformation of the transverse momenta in the ${\bf k}$-space resulting into the geometric phase $\gamma_{ph}$. 

The link between the topological charge and the orbital AM of LG-modes of light beams \cite{29} has been a subject of confusion in the literature; Berry and Liu \cite{87} unjustifiably term it misconception \cite{88}. A recent paper on the propagation of cyclone catastrophe
beams (CCB) investigates the relation between phase vortices and orbital AM \cite{89}. Of particular interest in connection with GP is that the superposition of LG-modes discussed in \cite{89} could be reformulated utilizing the contribution of the second part of $M^{\alpha \mu\nu}_{EM}$ in Eq.(57); the quantity $(A^\mu x^\nu -A^\nu x^\mu)$ in this expression contains $({\bf r} \times {\bf A})$ that has natural identification as AM in view of ${\bf r} \times ({\bf p} -\frac{e}{c}{\bf A})$.  We suggest this line of approach based on GP and AM holonomy to settle the issue of the relation between orbital AM and the charge of phase vortex in optics.

{\bf RB:~} The polarization state-space for light, the Poincare sphere, has been used to illustrate the Pancharatnam phase holonomy in the literature \cite{90} since the original work of Pancharatnam \cite{20}. Past few years have witnessed immense activities on the  class of polarized light beams that possess space-variant polarization nicely discussed in Born and Wolf \cite{11}. A beautiful application of Pancharatnam's idea to relate the phases of space-varying polarized light has been given by Nye \cite{91}. In fact, the subject of crystal dislocations and disclinations in liquid crystals as a paradigm is suggested for space-variant polarization ellipses by Nye; recent study on 2D liquid crystals \cite{92} may benefit from the insights obtained in \cite{91}. 

Briefly stated Nye's idea is to calculate the phase relation between space-varying polarization states of light at two nearby points in space assuming that the polarization ellipses differ in size, shape and orientation in space using Pancharatnam's theorem. The phase difference between two complex vector fields is $\arg ({\bf V}^*_1 \cdot {\bf V}_2)$ from the intensity 
\begin{equation}
I \propto ({\bf V}_1 +{\bf V}_2) \cdot ({\bf V}_1^* +{\bf V}_2^*)
\end{equation}
Using the expressions (20)-(21), and defining the normal vector ${\bf n} = {\bf a} \times {\bf b}$ the differential form of the phase is obtained to be
\begin{equation}
d \delta = -\frac{2 a b }{a^2+b^2} d \alpha +d \epsilon
\end{equation}
Here $a=|{\bf a}|,~b=|{\bf b}|$; $d \alpha$ is the rotation angle of the axes of the ellipse about the normal. Both $d \delta$ and $d \alpha$ are not integrable. Define ${\bf k}_\delta = \frac{\partial \delta}{\partial {\bf r}}, ~{\bf k}_\alpha = \frac{\partial \alpha}{\partial {\bf r}}$ then from Eq.(126) we get
\begin{equation}
{\bf k}_\delta = -\frac{2 a b }{a^2+b^2} {\bf k}_\alpha +{\bf \nabla} \epsilon
\end{equation}
Introducing amplitudes and phases analogous to the ones used in Eq.(60)) and Eq.(61) Nye finally arrives at the expression for ${\bf k}_\delta$ identical to the weighted mean Eq.(63). According to him this is the main result of the paper \cite{91}.

We have argued in \cite{40} that Nye used the Pancharatnam connection correctly but the associated phase for the vector vortex light beams is not Pancharatnam phase: the spatially evolving spin redirection phase is intrinsically involved. This argument needs elaboration and rigour. Pancharatnam \cite{20} considers the general question of the properties of two polarized light beams propagating in the same direction and puts forward the proposition: "The intensity I of the beam obtained on combining two mutually coherent beams 1 and 2, of intensities $I_1$ and $I_2$ in the states of polarization A and B respectively, will be given by the general interference formula
\begin{equation}
I= I_1 +I_2 + 2 \sqrt{I_1I_2} \cos \frac{1}{2}c \cos \delta
\end{equation}
Here $c$ is the angular separation of the states A and B on the Poincare sphere."  The geometric phase $\delta$  is equal to half of the area of triangle $C^\prime BA$ where $C^\prime$ is the diametrically opposite point to the state of the polarization of the resultant beam $C$. 

Now, three points on the Nye's paper emerge immediately: (i) Eq.(125) represents Pancharatnam's Eq.(128) using complex vector fields, (ii) a generalization to the two light beams propagating in different directions is made by Nye compared to the fixed direction of propagation of two light beams represented on the Poincare sphere in \cite{20}, and (iii) the polarization states on the Poincare sphere, for example, $(A, B, C, C^\prime )$ in Pancharatnam's paper are not space-varying whereas in Nye's case the spatially varying polarization ellipses are considered. As a consequence of these points the geometric phase appearing in Eq.(125) cannot be similar to the one appearing in Pancharatnam's Eq.(128).   
Eq(126) and Eq.(127) are useful to classify singular nature of the vector vortices as the weighted mean of ${\bf \nabla}\epsilon$ and ${\bf k}_\alpha$ is regular \cite{91}. However, the drawback of these definitions is that the meaning pf  GP becomes obscure and the contributions to GP arising from the polarization changes and the propagation vector changes for a general vector vortex are not discernible. 

Nye's theory as well as the physics of space-varying polarization need modifications and reinterpretations. The proposition made in \cite{40} based on spatially evolving spin redirection phase seeks AM holonomy manifestations in the backscattered polarization experiment \cite{93}, q-plate experiment on the claimed spin to orbital AM conversion \cite{94} , and the experiment on the tightly focused beams \cite{95}. It would be interesting to investigate  GP of non-geodesic circles \cite{90} and relate it with AM holonomy; we suggest the kind of experiment reported recently \cite{96} could be re-designed to explore this issue.

A general theory for vector vortex light beams that has well-defined propagation vector field and polarization vector field is almost impossible \cite{11, 64, 65, 91}; here we use fields because constant propagation vector and polarization vector for plane waves have to be generalized to the functions of space-time in general. A useful approach could be to develop an algorithm that separates relative importance of spatial modes, polarization modes, and polarization singularities to calculate the GP arising from propagation vector changes (for scalar waves), Pancharatnam phase, and spin redirection phase as approximations. For example, the HG-modes or LG-modes with a definite state of polarization would have only GP given by Eq.(118) and Eq.(119). Now, superposition of the spatial modes with different polarizations results into the vector beams, for example, radially and azimuthally polarized beams. The experimental results on such beams generated by space-variant gratings \cite{97} could be re-analyzed based on the proposed algorithm; note that the statements in \cite{97} invoking the Pancharatnam phase for the interpretation are not justified as it is the GP in wave vector space that is natural for these beams. The algorithm may be supplemented with the idea of a physical Poincare sphere \cite{98} to get new physical insights in these situations. 

\section{\bf Conclusion}

Geometric phases in optics have nice mathematical and experimental renditions in the vast literature on this subject \cite{1, 8, 24, 25, 26}. The physical origin of GP in optics in terms of the angular momentum exchange was proposed in 1992 \cite{2} along with a conjecture on AM holonomy. A confusion prevails in the literature on the meaning of the suggested AM transfer and the AM holonomy conjecture. The present contribution delineates the issues involved in the confusions/misinterpretations, and resolves them gaining new physical insights into the nature of the light waves and GP in optics. The proposition of the AM holonomy may throw light on the physics of structured light beams of current interest.

GP for a single photon and spin to orbital AM conversion for a single photon are intricate issues as we have argued in Section III that these correspond to single-photon states, not a real physical photon that has intrinsic spin of topological origin having the attributes of a particle \cite{36, 37}. Pancharatnam phase for a single photon was critically analyzed \cite{99} that was reported by Kwiat and Chiao \cite{100}. The question is to incorporate the idea of a topological photon in a theory that leads to the Maxwell EM field theory for large number of photons. The statements in the literature using the word "photon", for example, the orbital AM per photon or spin of a linearly polarized, RCP or LCP photon are more like euphuism. Sometimes they do provide intuitive appeal, and help in fixing the physical concepts: authors in \cite{86} discussing orbital AM in the process of topological inversion make use of the word "number of photons" to fix the notion of AM conservation. Going beyond it, we need a genuine photon fluid theory that so far remains elusive. Statistical approach and incorporation of topology have to be integral in this endeavour. It seems photon dynamics in physical space \cite{101} critically examined in \cite{85} could be an important ingredient in developing macroscopic theory of an ensemble of topological photons. 

{ ~~\bf APPENDIX}

The notation for vectors and metric in 4D space-time is as follows. The metric is $\eta_{\mu\nu} =\eta^{\mu\nu} = diag (1, -1, -1, -1)$. Four vectors are $A^\mu = A^0, {\bf A}$. Partial derivatives are given by
\begin{equation}
\partial_\mu= \frac{\partial }{\partial x^\mu} =( \frac{\partial }{\partial ct}, {\bf \nabla})
\end{equation}

\end{document}